\documentclass[twocolumn,aps,prl,superscriptaddress,nofootinbib,showpacs,preprintnumbers]{revtex4-1}%
\usepackage{graphicx}

\usepackage{amsmath}
\usepackage{slashed}
\usepackage{amssymb}
\usepackage{bm}
\usepackage{color}

\usepackage[hypertex]{hyperref}

\begin{document}


\preprint{DESY~19--111\hspace{14.cm} ISSN 0418--9833$\hphantom{XXXXX}$}
\preprint{June 2019\hspace{18.4cm}}


\boldmath
\title{Double prompt $J/\psi$ hadroproduction in the parton Reggeization
  approach with high-energy resummation}
  
\unboldmath

\author{Zhi-Guo He}
\affiliation{{II.} Institut f\"ur Theoretische Physik, Universit\"at Hamburg,
Luruper Chaussee 149, 22761 Hamburg, Germany}
\author{Bernd A. Kniehl}
\affiliation{{II.} Institut f\"ur Theoretische Physik, Universit\"at Hamburg,
Luruper Chaussee 149, 22761 Hamburg, Germany}
\author{Maxim A. Nefedov}
\affiliation{{II.} Institut f\"ur Theoretische Physik, Universit\"at Hamburg,
Luruper Chaussee 149, 22761 Hamburg, Germany}
\affiliation{Samara National Research University, Moskovskoe Shosse 34, 443086 
Samara, Russia}
\author{Vladimir A. Saleev}
\affiliation{Samara National Research University, Moskovskoe Shosse 34, 443086 
Samara, Russia}

\date{\today}

\begin{abstract}
  We study double prompt $J/\psi$ hadroproduction within the nonrelativistic-QCD  factorization formalism adopting the parton Reggeization approach to
  treat initial-state radiation in a gauge invariant and infrared-safe way.
  We present first predictions for the cross section distributions in the
  transverse momenta of the subleading $J/\psi$ meson and the $J/\psi$ pair.
  Already at leading order in $\alpha_s$, these predictions as well as those
  for the total cross section and its distributions in the invariant mass
  $m_{\psi\psi}$ and the rapidity separation $|Y|$ of the $J/\psi$ pair nicely
  agree with recent ATLAS and CMS measurements, except for the
  large-$m_{\psi\psi}$ and large-$|Y|$ regions, where the predictions
  substantially undershoot the data.
  In the latter regions, BFKL resummation is shown to enhance the cross
  sections by up to a factor of two and so to improve the description of the
  data.
\end{abstract}

\maketitle

Despite concerted experimental and theoretical endeavors ever since the
discovery of $J/\psi$ meson more than four decades ago, its production
mechanism has remained mysterious; for a recent review, see
Ref.~\cite{Brambilla:2010cs}.
The factorization approach \cite{Bodwin:1994jh} to nonrelativistic QCD (NRQCD)
\cite{Caswell:1985ui} endowed with velocity scaling rules \cite{Lepage:1992tx}
for the long-distance matrix elements (LDMEs), which is by far the most
acceptable candidate theory for heavy-quarkonium production and decay and has
been elaborated at next-to-leading order (NLO), has been challenged by the
long-standing $J/\psi$ polarization puzzle \cite{Butenschoen:2012px} and by the
inadequate description of $\eta_c$ hadroproduction data with $J/\psi$ LDMEs
converted via heavy-quark spin symmetry \cite{Butenschoen:2014dra}.
The high flux of incoming partons at LHC allows us to study $J/\psi$
production more thoroughly, also in association with other charmonia,
bottomonia, $W$ or $Z$ bosons, so as to pin down the $J/\psi$ production
mechanism.
Among these production processes, double $J/\psi$ hadroproduction is of special
interest because $J/\psi$ formation takes place there twice, making this
particularly sensitive to the nonperturbative aspects of NRQCD
\cite{Barger:1995vx}.
Moreover, this is believed to be an exquisite laboratory to study double parton
scattering (DPS) and to extract its key parameter, the effective cross section
$\sigma_{\mathrm{eff}}$ \cite{Kom:2011bd}.

In recent years, double prompt $J/\psi$ hadroproduction has been measured
extensively by the LHCb~\cite{Aaij:2011yc}, CMS~\cite{Khachatryan:2014iia}, and
ATLAS~\cite{Aaboud:2016fzt} Collaborations at the CERN LHC, and by the D0
Collaboration~\cite{Abazov:2014qba} at the FNAL Tevatron. 
On the theoretical side, the complete NRQCD results at leading order (LO) have
been obtained recently \cite{He:2015qya}.
For some channels, the relativistic corrections \cite{Li:2013csa} and NLO QCD
corrections \cite{Sun:2014gca} are also available.
The available NRQCD predictions can explain the LHCb and D0 data, and to some
extent also the CMS data, reasonably well.
However, these single-parton scattering (SPS) predictions only amount to a few
percent of the CMS data in the regions of large invariant mass $m_{\psi\psi}$ or
rapidity separation $|Y|=|y_1-y_2|$ of the $J/\psi$ pair, although the
color-octet contributions, in particular those involving $t$-channel gluon
exchanges, enhance the QCD-corrected \cite{Sun:2014gca} color-singlet
contribution of $gg\to2c\bar{c}({}^3S_1^{[1]})$ there by more than one order of
magnitude.
The value of $\sigma_{\mathrm{eff}}$ extracted by fitting the DPS contribution on
top of this \cite{Abazov:2014qba} is considerably smaller than typical values
from other processes, and the resulting SPS plus DPS results still undershoot
the CMS data in the upper $m_{\psi\psi}$ and $|Y|$ bins \cite{Lansberg:2014swa}.

As noticed in Ref.~\cite{He:2015qya}, the CMS kinematic conditions
\cite{Khachatryan:2014iia} render $2\to 3$ subprocesses predominant, which
enter at NLO in the collinear parton model (CPM). 
However, a complete NLO NRQCD computation is presently out of reach from the
technical point of view.
On the conceptual side, the conventional NRQCD factorization formalism needs
to be extended to cope with the double $P$-wave case \cite{He:2018hwb}.
Moreover, the perturbative expansion is spoiled for small values of the
$J/\psi$ pair transverse momentum $p_T^{\psi\psi}$.
The characteristic scale $\mu\sim[(4 m_c)^2+(p_T^{\psi\psi})^2]^{1/2}$ of the
hard-scattering processes of double $J/\psi$ hadroproduction satisfies
$\Lambda_{\mathrm{QCD}}\ll\mu\ll\sqrt{S}$, where
$\Lambda_{\mathrm{QCD}}$ is the asymptotic scale parameter and
$\sqrt{S}$ is the center-of-mass energy.
We are thus accessing the high-energy Regge regime, where the NLO QCD
corrections can be largely accounted for through the unintegrated  parton
distribution functions (unPDFs) in the parton Reggeization approach (PRA)
\cite{Kniehl:2014qva} based on Lipatov's effective field theory formulated with
the non-Abelian gauge invariant action \cite{Lipatov:1995}.
The PRA has already been successfully applied to the interpretation of
measurements of single heavy-quarkonium hadroproduction
\cite{Kniehl:2006sk,Kniehl:2006vm}.
In the large-$|Y|$ region, the two $J/\psi$ mesons are well separated obeying
multi-Regge kinematics (MRK).
For subprocesses containing $t$-channel gluon exchange type diagrams,
there will be large logarithms of form $(\alpha_s\ln|s/t|)^n$ in the
higher-order QCD corrections, where $s$ and $t$ are the Mandelstam variables of
the partonic $2\to2$ Born process.
Such large logarithms can be resummed by the Balitsky-Fadin-Kuraev-Lipatov
(BFKL) formalism \cite{Kuraev:1976ge}.
Recently, BFKL resummation has been studied for single $J/\psi$
\cite{Kotko:2019kma} and $J/\psi$ plus jet \cite{Boussarie:2017oae} inclusive
hadroproduction.
In this Letter, we will take a crucial step towards a full-fledged NLO NRQCD
study of double prompt $J/\psi$ hadroproduction, by adopting the PRA and
performing BFKL resummation.

Owing to the PRA and NRQCD factorization, the cross section of inclusive
double prompt $J/\psi$ hadroproduction can be expressed as: 
\begin{eqnarray}\label{PRA+NRQCD}
  \lefteqn{d\sigma^{\mathrm{PRA}}(AB\to 2J/\psi+X)
    =\sum_{m,n,H_1,H_2}\int\frac{dx_1}{x_1}
    \int\frac{d^2 \mathbf{k}_{1T}}{\pi}} \nonumber\\
&&{}\times\int\frac{dx_2}{x_2}\int\frac{d^2\mathbf{k}_{2T}}{\pi}
\Phi_{R^+/A}(x_1,t_1,\mu^2)\Phi_{R^-/B}(x_2,t_2,\mu^2)\nonumber\\
&&{}\times d\hat{\sigma}_{mn}^{\mathrm{PRA}}\langle\overline{\mathcal{O}}^{H_1}(m)
\rangle\langle\overline{\mathcal{O}}^{H_2}(n)\rangle,
\end{eqnarray}
where $d\hat{\sigma}_{mn}^{\mathrm{PRA}}$ is the short-distance coefficient (SDC)
of the partonic subprocess $R^{+}(k_1)R^{-}(k_2)\to c\bar{c}(m)c\bar{c}(n)+X$,
$\Phi_{R^\pm/A,B}(x_{1,2},t_{1,2},\mu^2)$ are the unPDFs of the Reggeized gluons
$R^{\pm}$ with four-momenta $k_{1,2}^\mu=x_{1,2}K_{1,2}^\mu+k_{1,2T}^\mu$ and
virtualities $t_{1,2}=-k_{1,2}^2=\mathbf{k}_{1,2T}^2$, 
$K_{1,2}^\mu$ are the four-momenta of the colliding hadrons $A,B$ with
light-cone components $K_1^-=K_2^+=0$ ($K_i^\pm=K_i^0\pm K_i^3$), and
$\langle\overline{\mathcal{O}}^{H}(m)\rangle$ is the product of LDME
$\langle\mathcal{O}^{H}(m)\rangle$ of $H=J/\psi,\chi_{cJ},\psi^\prime$ and 
the branching fraction $\mathrm{BR}(H\to J/\psi+X)$, with the understanding
that $\mathrm{BR}(H\to J/\psi+X)=1$ if $H=J/\psi$.
Since partonic subprocesses initiated by Reggeized quarks and antiquarks are
greatly suppressed by their unPDFs, we may disregard them here.
Furthermore, we may neglect the $\chi_{c0}$ feed-down contribution because
$\mathrm{BR}(\chi_{c0}\to J/\psi+X)=1.40\%$ \cite{Tanabashi:2018kda} is so
small.

\begin{figure}
\centering
\begin{tabular}{c}
\includegraphics[width=0.9\linewidth]{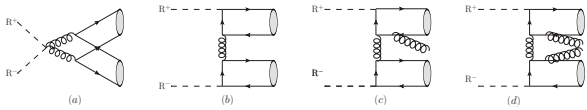}
\end{tabular}
\caption{\label{Feynman}%
  Typical Feynman diagrams for $R^{+}R^{-}\to c\bar{c}(m)c\bar{c}(n)$ of
  types (a) non-$t$-channel gluon exchange and $t$-channel gluon exchange at
  (b) LO, (c) NLO, and (d) NNLO in $\alpha_s$.}
\end{figure}

Representative Feynman diagrams for the partonic subprocess
$R^{+}R^{-}\to c\bar{c}(m)c\bar{c}(n)$ at LO in $\alpha_s$ are depicted in
Figs.~\ref{Feynman}(a) and (b).
By the Feynman rules of Ref.~\cite{Lipatov:1995}, they come in the same
topologies as those for $gg$ fusion in the CPM.
To discuss the BFKL resummation effect conveniently, we divide the partonic 
subprocesses into three categories, according to the order in $\alpha_s$ where
$t$-channel gluon exchanges emerge for the first time, namely
(i) LO $t$-channel (LT), with
$m,n={}^1\!S_0^{[8]},{}^3\!S_1^{[8]},{}^3\!P_J^{[1,8]}$;
(ii) NLO $t$-channel (NLT), with $m={}^3\!S_1^{[1]}$ and
$n={}^1\!S_0^{[8]},{}^3\!S_1^{[8]},{}^3\!P_J^{[1,8]}$; and
(iii) NNLO $t$-channel (NNLT), with $m,n={}^3\!S_1^{[1]}$;
see Figs.~\ref{Feynman}(b)--(d).

\begin{table*}
  \caption{\label{LDMEs}%
    Adopted values of LO NRQCD LDMEs in units of GeV${}^3$.}
\begin{tabular}{ccccccccc}
\hline
$\langle\mathcal{O}^{J/\psi}({}^3\!S_1^{[1]})\rangle$ &
$\langle\mathcal{O}^{J/\psi}({}^1\!S_0^{[8]})\rangle$ &
$\langle\mathcal{O}^{J/\psi}({}^3\!S_1^{[8]})\rangle$ &
$\langle\mathcal{O}^{\psi^\prime}({}^3\!S_1^{[1]})\rangle$ & 
$\langle\mathcal{O}^{\psi^\prime}({}^1\!S_0^{[8]})\rangle$ &
$\langle\mathcal{O}^{\psi^\prime}({}^3\!S_1^{[8]})\rangle$ & 
$\frac{\langle\mathcal{O}^{\chi_{c0}}({}^3\!P_0^{[1]})\rangle}{m_c^2}$ &
$\langle\mathcal{O}^{\chi_{c0}}({}^3\!S_1^{[8]})\rangle$&
$\frac{\langle\mathcal{O}^{J/\psi(\psi^\prime)}({}^3\!P_J^{[8]})\rangle}{m_c^2}$ \\
\hline
1.16 & $3.61\times10^{-2}$ & $1.25\times10^{-3}$ & $0.76$ & $2.19\times10^{-2}$ &
$3.41\times10^{-4}$ & $4.77\times10^{-2}$ & $5.29\times10^{-4}$ &0 \\
\hline
\end{tabular}
\end{table*}

We first compute the LO contributions to all the three categories.
Due to lack of space, we relegate the details of our calculation to a separate
paper.
In contrast to other $k_T$ factorization approaches \cite{Baranov:2015cle}, the
PRA yields gauge invariant SDCs with off-shell initial-state partons, which
provides a strong check for our analytic calculations.
In the collinear limits $t_{1,2}\to 0$, we recover the CPM formulas
\cite{He:2015qya}, which constitutes yet another nontrivial check.
In the numerical analysis, we adopt the Kimber-Martin-Ryskin scheme
\cite{Kimber:2001sc} to generate the unPDFs from the LO CPM PDFs of
Ref.~\cite{Martin:2009iq}, which come with $\alpha_s(M_Z)=0.13939$.
We choose the renormalization and factorization scales to be 
$\mu_r=\mu_f=\xi[(4m_c)^2+\bar{p}_T^2]^{1/2}$,
where $m_c=1.5~\mathrm{GeV}$,
$\bar{p}_T=(p_T^{H_1}+p_T^{H_2})/2$, and $\xi$ is varied between $1/2$ and 2 about
its default value 1 to estimate the scale uncertainty.
In the case of feed-down from $H=\chi_{c_1},\chi_{c_2},\psi^\prime$, we put
$p_T^{J/\psi}=p_T^{H}M_{J/\psi}/M_H$, which is a good approximation because
$p_T^{J/\psi}\gg M_{H}-M_{J/\psi}$ \cite{Ma:2010vd}.
For the $J/\psi$, $\chi_{cJ}$, and $\psi^\prime$ LDMEs, we use the values
specified in Table~\ref{LDMEs}.
The color-singlet results have been derived from the Buchm\"uller-Tye potential
in Ref.~\cite{Eichten:1995ch}.
The color-octet results have been fitted to LHC data of inclusive single
charmonium hadroproduction \cite{ATLAS:2014ala} in the very theoretical
framework described above; they supersede pre-LHC results \cite{Kniehl:2006sk}.
For $H=J/\psi,\psi^\prime$, there is a strong correlation between
$\langle\mathcal{O}^{H}({}^1\!S_0^{[8]})\rangle$ and
$\langle\mathcal{O}^{H}({}^3\!P_0^{[8]})\rangle$, so that only a linear
combination of them can be determined.
We may thus put $\langle\mathcal{O}^{H}({}^3\!P_0^{[8]})\rangle=0$.
We have checked that the theoretical uncertainties in our predictions for
double prompt $J/\psi$ hadroproduction due to this freedom are negligible.
All other input parameters are adopted from Ref.~\cite{Tanabashi:2018kda}.

\begin{figure*}
\centering
\begin{tabular}{ccc}
\includegraphics[width=0.33\linewidth]{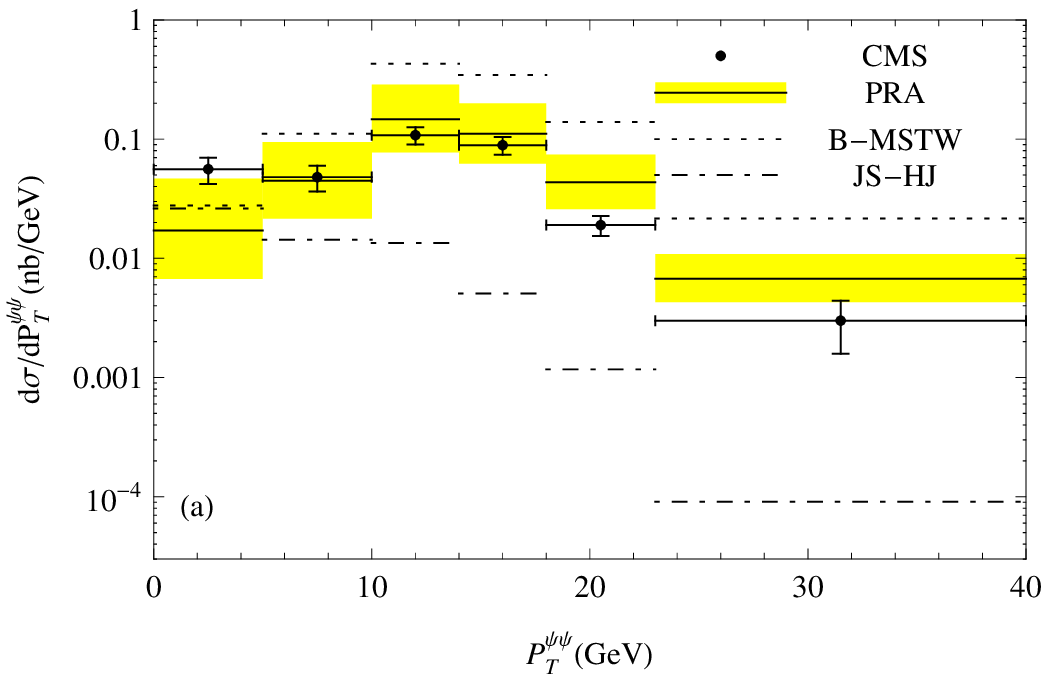}&
\includegraphics[width=0.33\linewidth]{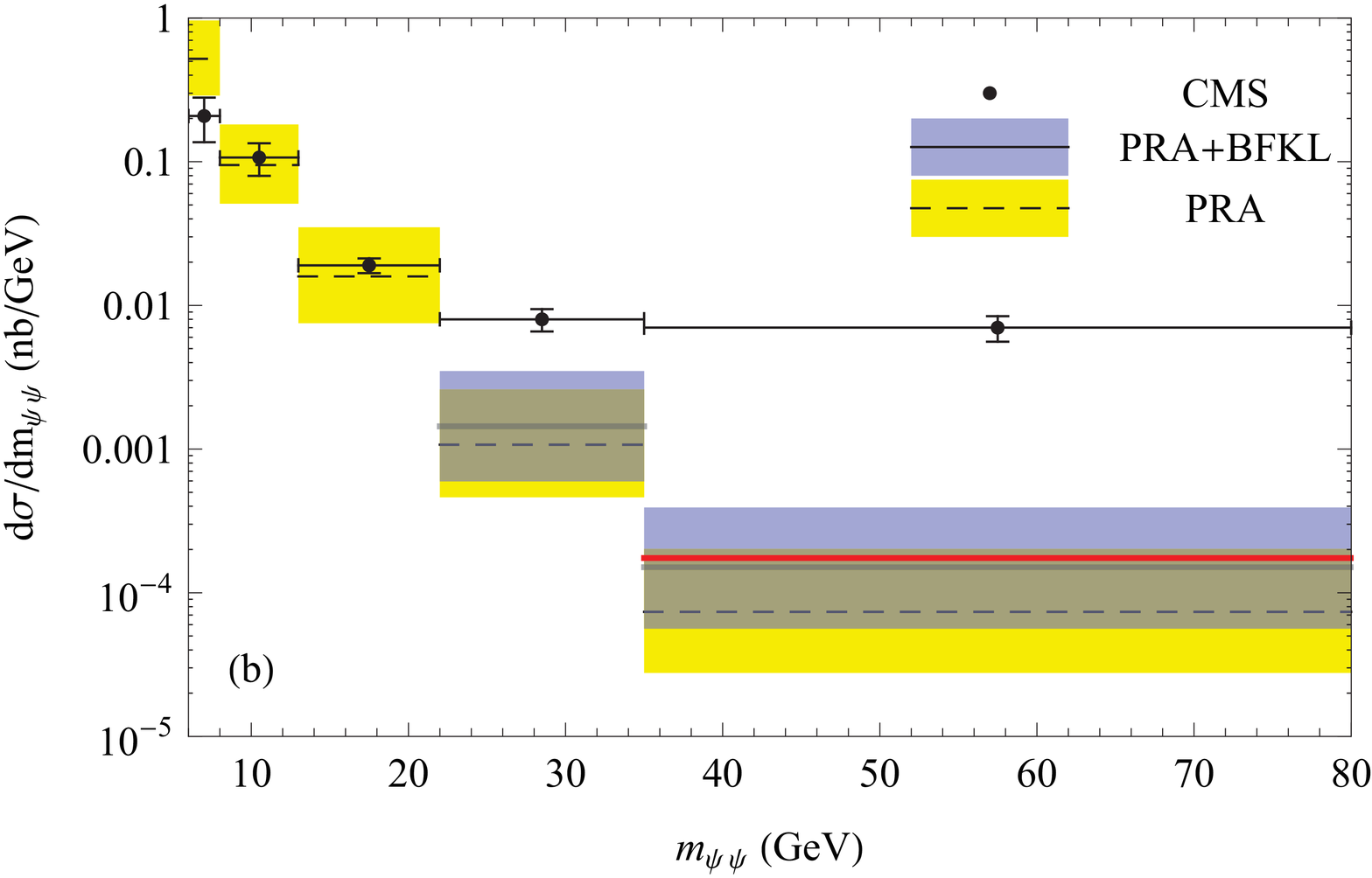}&
\includegraphics[width=0.33\linewidth]{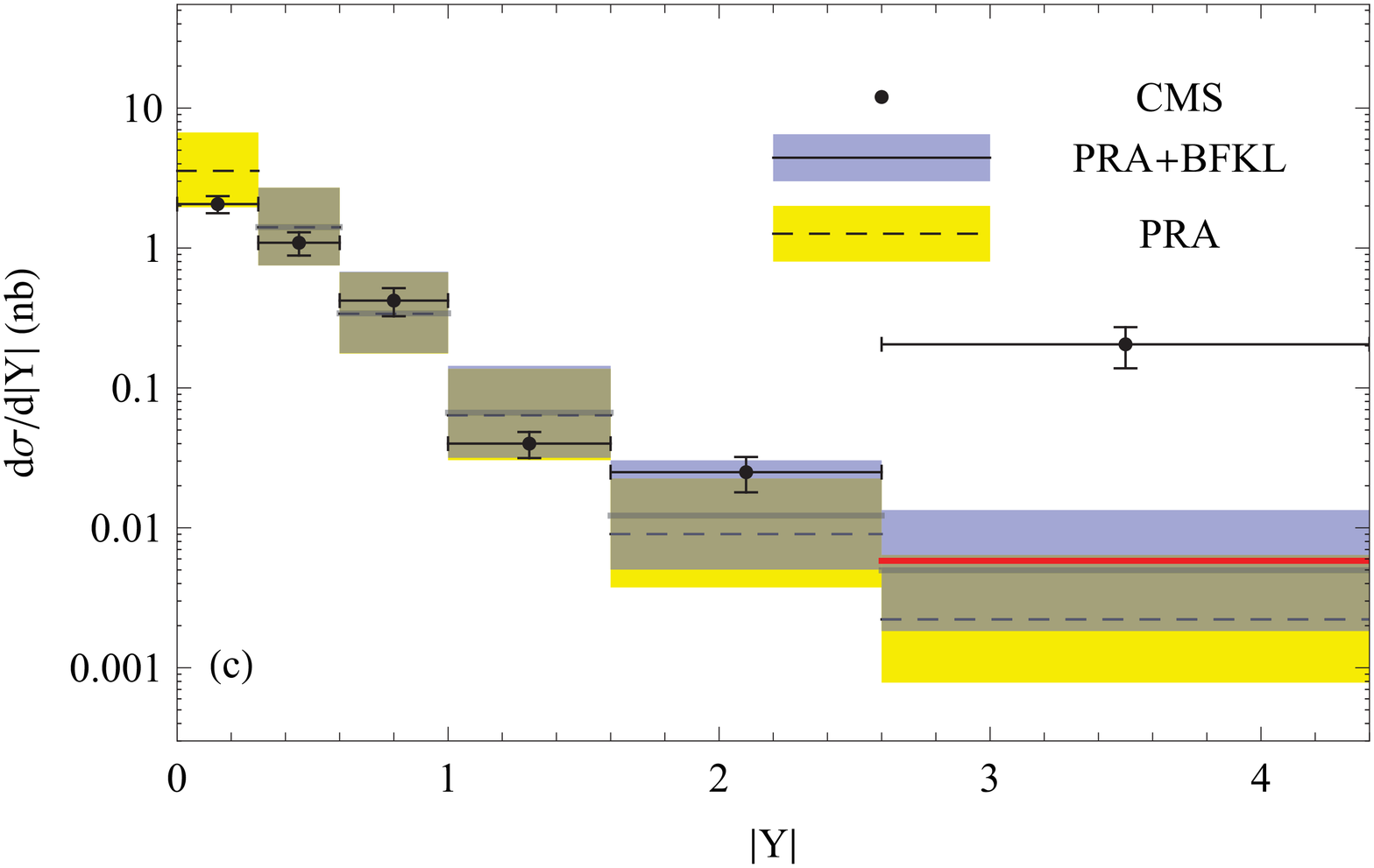}
\end{tabular}
\caption{\label{pracms}%
  The (a) $p_T^{\psi\psi}$, (b) $m_{\psi\psi}$, and (c) $|Y|$ distributions of
  double prompt $J/\psi$ production measured by CMS \cite{Khachatryan:2014iia}
  are compared to our LO NRQCD predictions in the PRA without (dashed lines)
  and with BFKL resummation (solid lines) including their scale uncertainties
  (yellow and blue bands).
  Adding the total ${\rm NLO}^{\ast}$ NLT contributions on top of the central
  LO NRQCD predictions in the PRA with BFKL resummation yields the red solid
  lines.
  Frame (a) also contains the evaluations with the unPDF sets of
  Refs.~\cite{Martin:2009iq,Blumlein:1995eu} (B-MSTW, dotted lines) and
  Ref.~\cite{Jung:2000hk} (JS-HJ, dot-dashed lines).}
\end{figure*}

The CMS data of prompt double $J/\psi$ hadroproduction were taken at
$\sqrt{S}=7~\mathrm{TeV}$ requiring for each $J/\psi$ meson to be in the
rapidity range $|y|<2.2$ and to satisfy a $y$-dependent minimum-$p_T$ cut, as
described in Eq.~(3.3) of Ref.~\cite{Khachatryan:2014iia}. 
The CMS total cross section
$\sigma_{\mathrm{CMS}}=(1.49\pm0.07\pm0.13)~\mathrm{nb}$ agrees with
our LO NRQCD prediction
$\sigma_{\mathrm{CMS}}^{\mathrm{PRA}}=1.68^{+1.32}_{-0.78}{}~\mathrm{nb}$ within
errors.
The CMS $p_T^{\psi\psi}$, $m_{\psi\psi}$, and $|Y|$ distributions are compared with
our predictions in Fig.~\ref{pracms}(a)--(c), respectively.
There is generally very good agreement, except for the upper two $m_{\psi\psi}$
bins and the upmost $|Y|$ bin, where the predictions undershoot the data.
The advancement of the PRA beyond the CPM is most striking for
Fig.~\ref{pracms}(a) because $p_T^{\psi\psi}=0$ at LO in the latter case, but it
is also significant for the $m_{\psi\psi}$ and $|Y|$ distributions, as may be
observed by comparing Figs.~\ref{pracms}(b) and (c) with their CPM
counterparts in Figs.~3 and 4 of Ref.~\cite{He:2015qya}.
In both cases, the predictions are substantially increased in the first
three bins, so as to nicely match the data, while the $K$ factors are of order
unity in the upmost bins.

The unPDF uncertainty may be assessed from Fig.~\ref{pracms}(a), which also
shows the evaluations using the set produced from our default PDF set
\cite{Martin:2009iq} as described in Ref.~\cite{Blumlein:1995eu} and the
set of Ref.~\cite{Jung:2000hk}.
The latter rapidly falls off with increasing $p_T^{\psi\psi}$ and significantly
undershoots the data in the upper $p_T^{\psi\psi}$ bins.
This is in line with Ref.~\cite{Maciula:2018bex}, where the unPDFs of
Ref.~\cite{Jung:2000hk} were found to yield a poor description of LHCb data of
single prompt $J/\psi$ production \cite{Aaij:2011jh} at large $p_T^{\psi}$.
This opens a novel perspective to constrain unPDFs.
To estimate the LDME uncertainty, we repeat the unresummed LO PRA evaluation in
the upmost $|Y|$ bin of Fig.~\ref{pracms}(c), which is most sensitive to the
color-octet LDMEs, using in turn the NLO CPM sets of
Refs.~\cite{Butenschoen:2011yh,Gong:2012ug}, albeit this is slightly
inconsistent.
Since Ref.~\cite{Butenschoen:2011yh} does not provide $\chi_{cJ}$ and
$\psi^\prime$ LDMEs, we use those of Ref.~\cite{Gong:2012ug} also here.
We thus find an enhancement by 57\% and a reduction by 7\% w.r.t.\ our default
result, respectively.
For a more detailed LDME analysis, see Ref.~\cite{Lansberg:2019fgm}.

ATLAS took their data at $\sqrt{S}=8~\mathrm{TeV}$ imposing the acceptance
cuts $p_T>8.5~\mathrm{GeV}$ and $|y|<2.1$ on each $J/\psi$ meson
\cite{Aaboud:2016fzt}.
They separately studied the central (I) and forward (II) $y$ regions of the
subleading $J/\psi$ meson ($J/\psi_2$), with $p_{2T}<p_{1T}$, namely $|y_2|<1.05$
and $1.05<|y_2|<2.1$.
Their respective total cross sections
$\sigma_{\mathrm{ATLAS,I}}=(82.2\pm8.3\pm6.3)~\mathrm{pb}$ and
$\sigma_{\mathrm{ATLAS,II}}=(78.3\pm9.2\pm6.6)~\mathrm{pb}$ are both compatible
with our LO PRA predictions
$\sigma_{\mathrm{ATLAS,I}}^{\mathrm{PRA}}=133.6^{+89.6}_{-52.2}{}~\mathrm{pb}$ and
$\sigma_{\mathrm{ATLAS,II}}^{\mathrm{PRA}}=105.2^{+73.8}_{-41.6}{}~\mathrm{pb}$.
Their respective $p_{2T}$, $p_T^{\psi\psi}$, and $m_{\psi\psi}$ distributions
are compared with our LO PRA predictions in Fig.~\ref{praatlas}.
We find fairly good agreement for the $p_{2T}$ and $p_T^{\psi\psi}$ distributions,
especially in region~II, with regard to both normalization and line shape.
In particular, the predictions faithfully reproduce the peaks of the measured
$p_T^{\psi\psi}$ distributions.
As for the $m_{\psi\psi}$ distributions, there is decent agreement for
$m_{\psi\psi}\alt40~\mathrm{GeV}$, while the predictions significantly undershoot
the data in the upper $m_{\psi\psi}$ bins, as in the CMS case above.

Also the LHCb \cite{Aaij:2011yc} and D0 \cite{Abazov:2014qba} measurements
reasonably agree with our PRA predictions.
The LHCb \cite{Aaij:2011yc} data at $\sqrt{S}=7~\mathrm{TeV}$, with acceptance
cuts $p_T<10~\mathrm{GeV}$ and $2<y<4.5$ on each $J/\psi$ meson, yield
$\chi^2/\mathrm{d.o.f.}=12.8/5$ with respect to our LO PRA prediction with
$\xi=1$ in the perturbatively safe region $m_{\psi\psi}>9~\mathrm{GeV}$
\cite{He:2015qya}, with experimental and theoretical errors overlapping. 
The fiducial cross section
$\sigma_\mathrm{SPS}=70\pm6(\mathrm{stat})\pm 22(\mathrm{syst})~\mathrm{fb}$
determined by D0 \cite{Abazov:2014qba} at $\sqrt{S}=7~\mathrm{TeV}$, with cuts
$p_T<10~\mathrm{GeV}$ and $2<y<4.5$, nicely agrees with our central LO PRA
prediction $\sigma_\mathrm{SPS}^\mathrm{PRA}=81.1~\mathrm{fb}$, where we have
included the reduction factor due the acceptance cuts on the decay muons
\cite{Abazov:2014qba} determined in Ref.~\cite{Qiao:2012wc}.

Although the LO CPM relationship
$m_{\psi\psi}=2[(2m_c)^2+(p_T^{J/\psi})^2]^{1/2}\cosh(|Y|/2)$ \cite{He:2015qya}
is evaded by the PRA, detailed analysis of the $m_{\psi\psi}$ and $|Y|$
distributions reveals that they are strongly correlated in the
large-$m_{\psi\psi}$ and -$|Y|$ regions.
As expected, the LT contributions greatly dominate there.
Specifically, they make up about 90\% or more of the total PRA predictions in
the upmost $m_{\psi\psi}$ bin in Fig.~\ref{pracms}(b), the upmost $|Y|$ bin in
Figs.~\ref{pracms}(c), the upmost $m_{\psi\psi}$ bin in Fig.~\ref{praatlas}(c),
and the upper four $m_{\psi\psi}$ bins in Fig.~\ref{praatlas}(f).
$t$-channel gluon exchanges, which appear in the LT contributions already at
LO, generate large logarithmic corrections of the type $(\alpha_s\ln|s/t|)^n$
in higher orders, which can be efficiently included via BFKL resummation.

\begin{figure}
\centering
\includegraphics[width=0.5\linewidth]{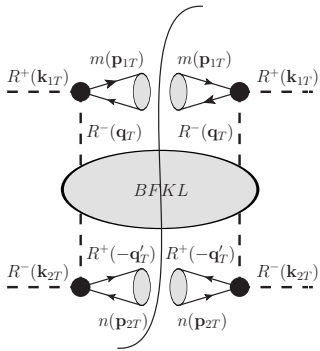}
\caption{\label{bfkldiagram}%
Schematic representation of Eq.~(\ref{resum1}).}
\end{figure}

BFKL resummation in the leading-logarithmic (LL) approximation is implemented
by replacing the SDCs in Eq.~(\ref{PRA+NRQCD}) with
\begin{eqnarray}\label{resum1}
  \lefteqn{\frac{d \hat{\sigma}^{\mathrm{BFKL}}_{mn}}
    {d\mathbf{p}_{1T}dy_1d\mathbf{p}_{2T}dy_2}=
\int \frac{d^2 \mathbf{q}_{T} d^2 \mathbf{q}^{\prime}_{T}}{(4\pi)^2S^2 x_1 x_2}}
\nonumber\\
&&{}\times\Psi^{(m)}_{+}(\mathbf{k}_{1T},\mathbf{p}_{1T},y_1)
G(\mathbf{q}_{T},\mathbf{q}^{\prime}_{T},Y)
\Psi^{(n)}_{-}(\mathbf{k}_{2T},\mathbf{p}_{2T},y_2),\hspace{0.5cm}
\end{eqnarray}
where $\Psi_{+}^{(m)}(\mathbf{k}_{1T},\mathbf{p}_{1T},y_1)$ and
$\Psi_{-}^{(n)}(\mathbf{k}_{2T},\mathbf{p}_{2T},y_2)$ are the impact factors
describing the partonic subprocesses 
$R^+(\mathbf{k}_{1T})R^-(\mathbf{q}_T)\to [c\bar{c}(m)](\mathbf{p}_{1T})$ and
$R^-(\mathbf{k}_{2T})R^+(-\mathbf{q}_T^\prime)\to [c\bar{c}(n)](\mathbf{p}_{2T})$,
obtained from the appropriate $2\to 1$ PRA matrix elements in
Ref.~\cite{Kniehl:2006sk} as explained in Ref.~\cite{Kovchegov:2012mbw}, and
$G(\mathbf{q}_{T},\mathbf{q}^{\prime}_{T},Y)$ is the BFKL Green function given by
Eq.~(3.80) in Ref.~\cite{Kovchegov:2012mbw}, generated from the initial
condition
$G(\mathbf{q}_{T},\mathbf{q}^{\prime}_{T},0)=\delta^{(2)}(\mathbf{q}_{T}
-\mathbf{q}^{\prime}_{T})$
via LL BFKL evolution in $Y$ \cite{Kuraev:1976ge}; see Fig.~\ref{bfkldiagram}
for a schematic representation of Eq.~(\ref{resum1}).
The resulting hadronic cross section is denoted as $d\sigma^{\mathrm{BFKL}}$.

$G(\mathbf{q}_{T},\mathbf{q}^{\prime}_{T},Y)$ depends exponentially on
$\alpha_s(\mu^2)$, which may produce a potentially large theoretical
uncertainty.
Several approaches have been proposed to remedy this.
As frequently done \cite{Ducloue:2013bva}, we adopt here the one
\cite{Brodsky:1998kn} based on a non-Abelian physical renormalization scheme
choice in connection with Brodsky-Lepage-Mackenzie optimal scale setting
\cite{Brodsky:1982gc}.
There remains a reference scale $\mu_0$.
We choose $\mu_0=\xi[|\mathbf{k}_{1T}||\mathbf{k}_{2T}|]^{1/2}$ and vary $\xi$
from $1/2$ to 2 about its default value 1 to estimate the residual scale
uncertainty in $G(\mathbf{q}_{T},\mathbf{q}^{\prime}_{T},Y)$.

We merge the full LO PRA calculation $d\sigma^{\mathrm{PRA}}$, appropriate in the
small-$|Y|$ region, and the LL-resummed LT contribution $d\sigma^{\mathrm{BFKL}}$,
appropriate in the large-$|Y|$ region, as
\begin{equation}\label{tot}
  d\sigma^{\mathrm{PRA+BFKL}}=d\sigma^{\mathrm{PRA}}+d\sigma^{\mathrm{BFKL}}
  -d\sigma^{\mathrm{BFKL},0},
\end{equation}
where the asymptotic term $d\sigma^{\mathrm{BFKL},0}$, which is obtained from
$d\sigma^{\mathrm{BFKL}}$ by replacing $G(\mathbf{q}_{T},\mathbf{q}^{\prime}_{T},Y)$
with $G(\mathbf{q}_{T},\mathbf{q}^{\prime}_{T},0)$, is to avoid double counting.
Equation~(\ref{tot}) smoothly interpolates from $d\sigma^{\mathrm{PRA}}$ at small
$|Y|$ values to $d\sigma^{\mathrm{BFKL}}$ at large $|Y|$ values.

The BFKL-improved PRA predictions thus evaluated are also included in
Figs.~\ref{pracms} and \ref{praatlas}.
Their uncertainties are obtained by combining the PRA and BFKL ones in
quadrature.
They exceed the PRA uncertainties only moderately, which indicates that the
scale uncertainties in $G(\mathbf{q}_{T},\mathbf{q}^{\prime}_{T},Y)$ are well
under control.
In the $p_{2T}$ and $p_T^{\psi\psi}$ distributions and in the lowest few bins
of the $m_{\psi\psi}$ and $|Y|$ distributions, the BFKL resummation effects are
so insignificant that we refrain from displaying the BFKL-improved results.
On the other hand, these effects are significant in the upper
$m_{\psi\psi}$ and $|Y|$ bins, where they may even double the pure PRA results,
so as to reduce the shortfall with respect to the CMS
\cite{Khachatryan:2014iia} and ATLAS \cite{Aaboud:2016fzt} data.
In the latter case, even agreement is reached in some medium $m_{\psi\psi}$ bins.
However, large gaps remain in the upmost $|Y|$ bin of CMS and the upmost few
$m_{\psi\psi}$ bins of CMS and ATLAS, to be explained by DPS.
Contrary to na\"{\i}ve expectations, the optimal scale turns out to be larger
than $\mu_0$, so that using the latter instead leads to an enhancement of the
BFKL-improved results, by factors of 1.0, 1.0, 1.1, 1.5, and 4.1 in the second
to sixth $|Y|$ bin of CMS.

\begin{figure*}
\centering
\begin{tabular}{ccc}
\includegraphics[width=0.33\linewidth]{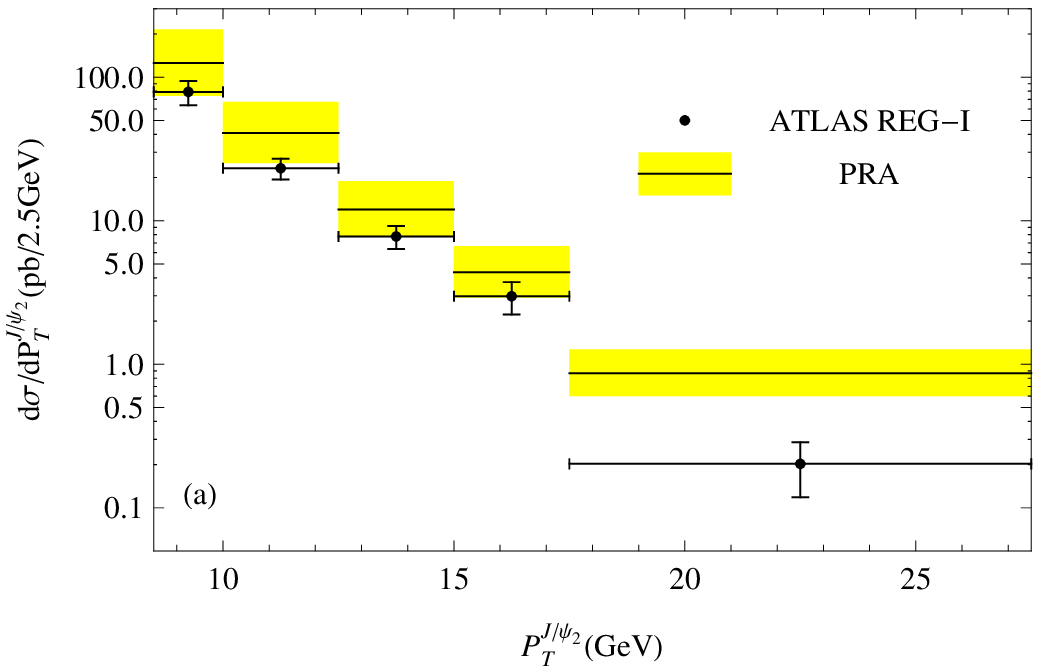}&
\includegraphics[width=0.33\linewidth]{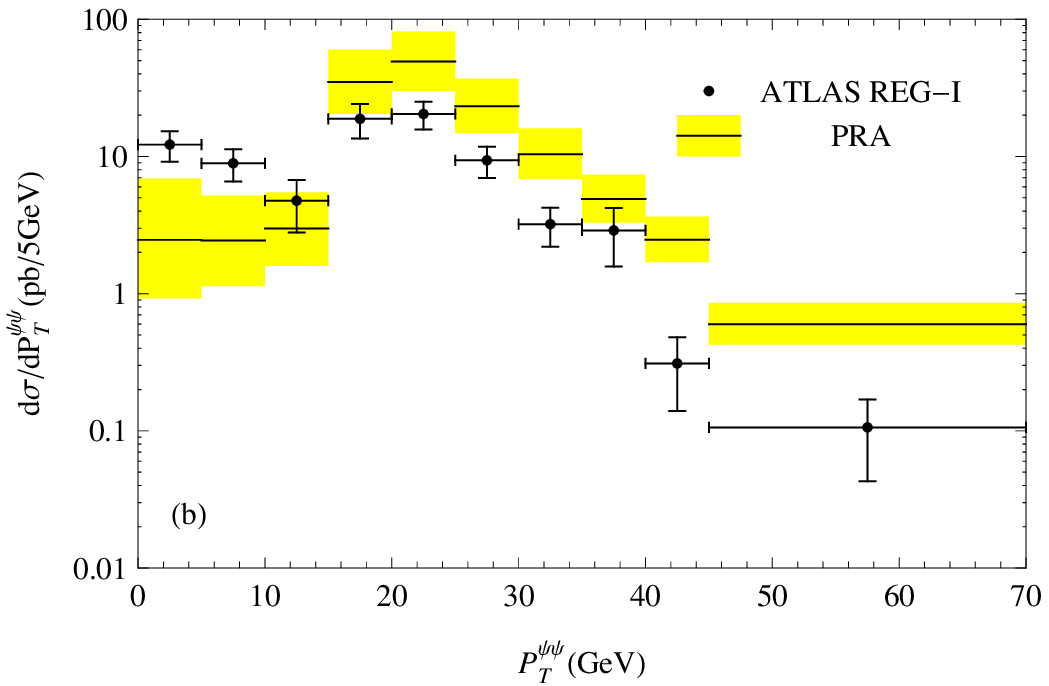}&
\includegraphics[width=0.33\linewidth]{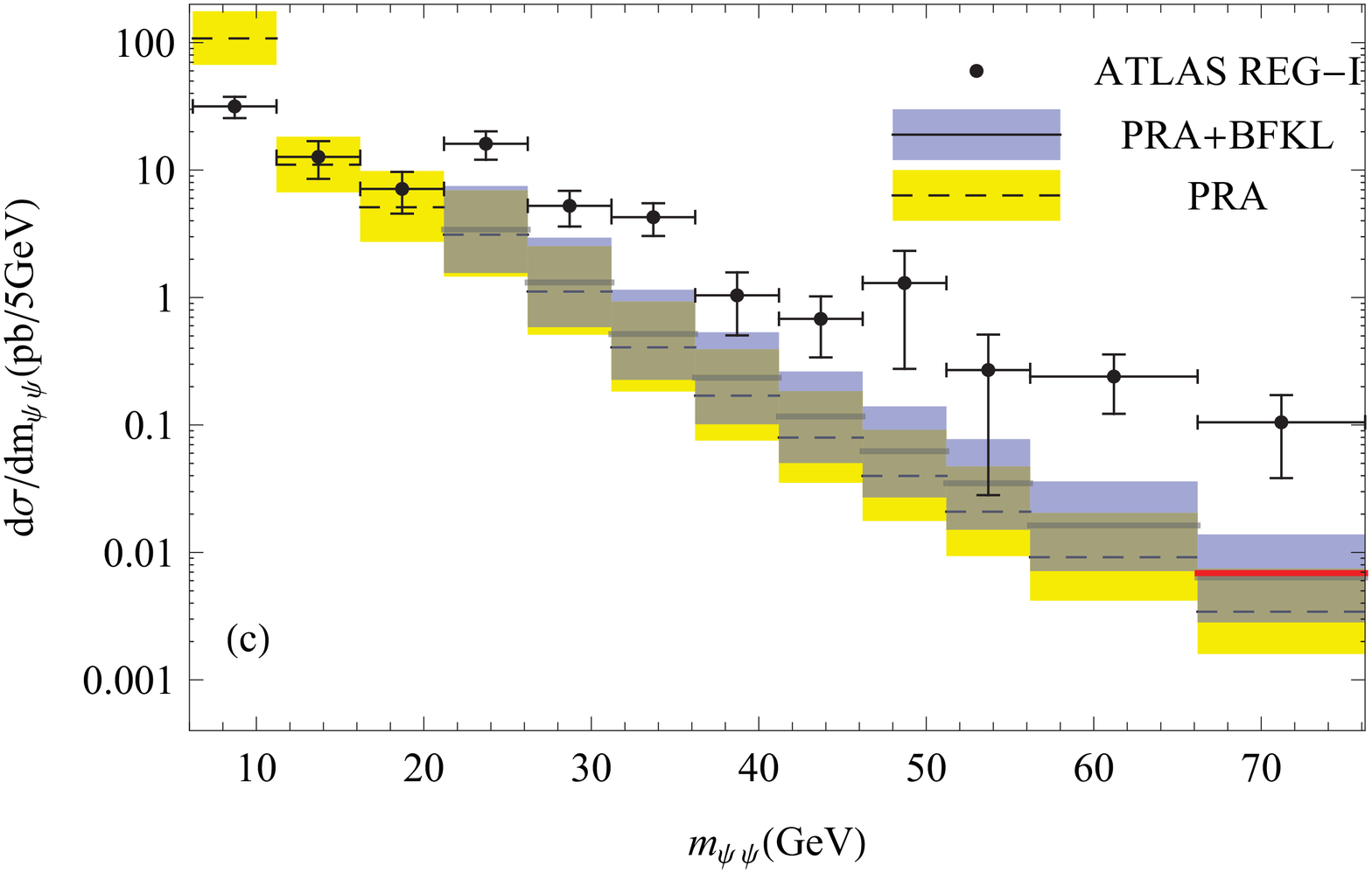}\\
\includegraphics[width=0.33\linewidth]{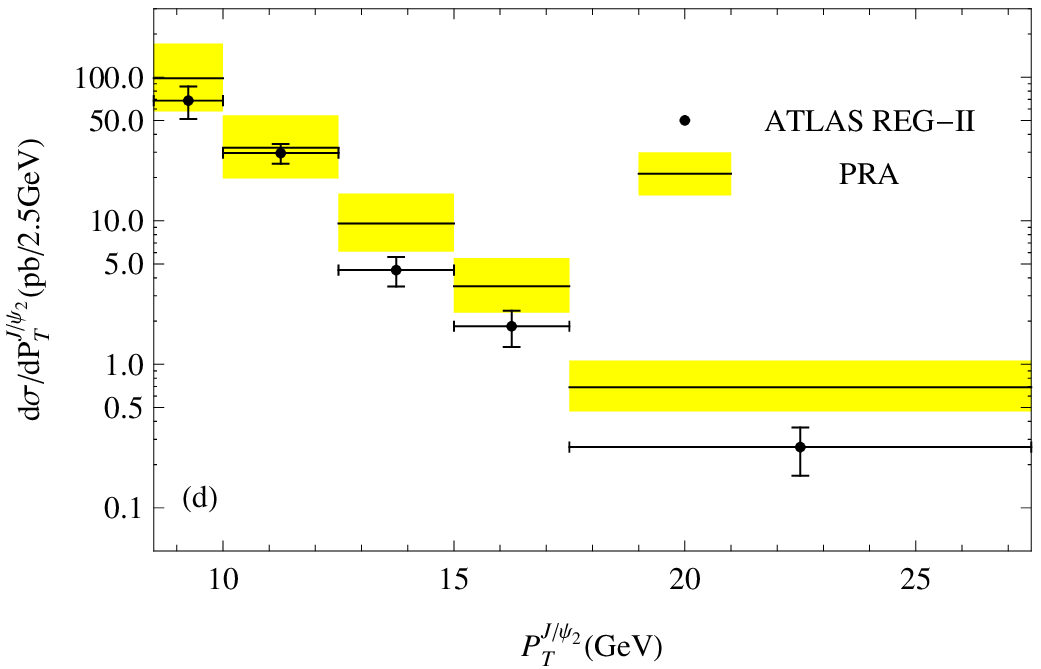}&
\includegraphics[width=0.33\linewidth]{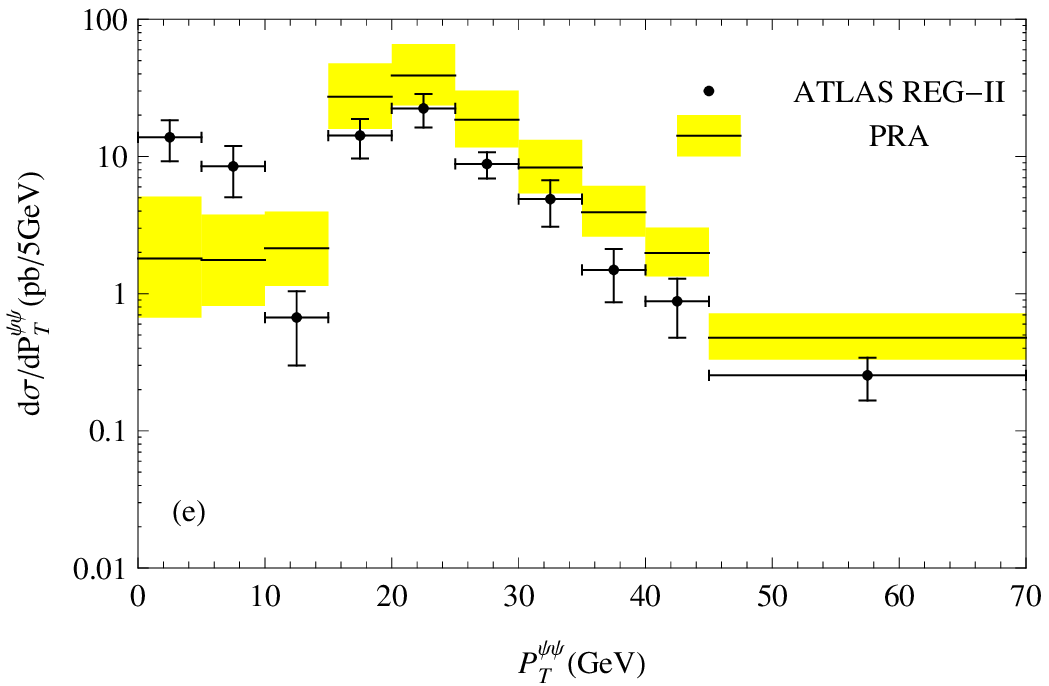}&
\includegraphics[width=0.33\linewidth]{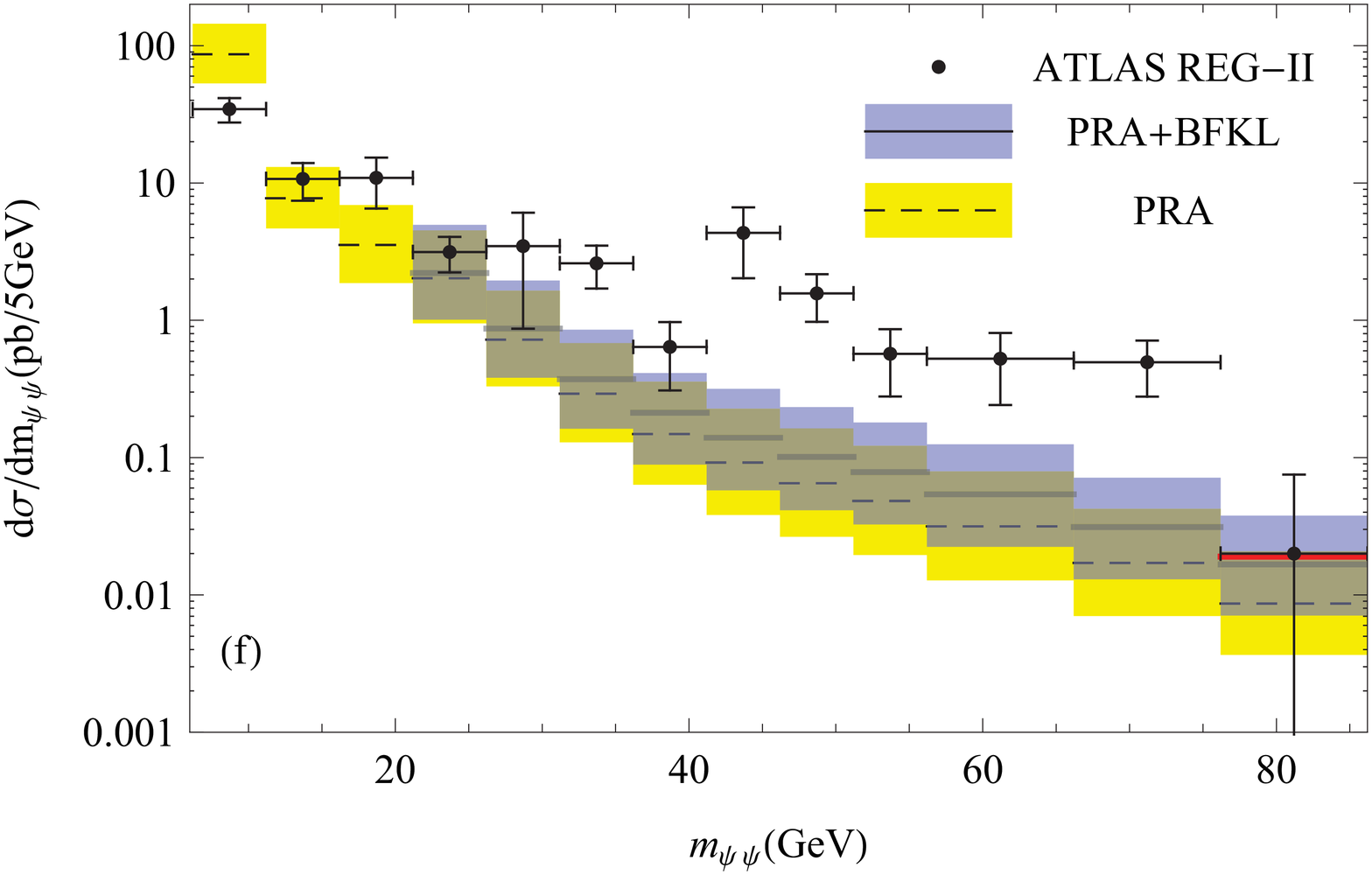}
\end{tabular}
\caption{\label{praatlas}%
As in Fig.~\ref{pracms}, but for the $p_{2T}$ (left column), $p_T^{\psi\psi}$
(middle column), and $m_{\psi\psi}$ (right column) distributions measured by
ATLAS \cite{Aaboud:2016fzt} in regions I (upper row) and II (lower row).}
\end{figure*}

Quantitative extractions of DPS contributions from the remaining discrepancies
are likely to be meaningful only after the complete NLO NRQCD corrections are
available, for the following reasons.
Firstly, conventional NRQCD factorization is known to break down at NLO for
double $P$-wave channels \cite{He:2018hwb}.
The quantitative influence of this is presently unclear.
Secondly, the NLO NRQCD corrections to the NLT subprocesses can be quite
sizable because their $\mathcal{O}(\alpha_s)$ suppression is expected to be
compensated by the relatively large values of the color-singlet LDMEs
$\langle\mathcal{O}^{J/\psi}({}^3\!S_1^{[1]})\rangle$ and
$\langle\mathcal{O}^{\psi^{\prime}}({}^3\!S_1^{[1]})\rangle$.
For the 
${}^3\!S_1^{[1]}+{}^1\!S_0^{[8]}/{}^3\!P_{1,2}^{[1]}$
channels, the type of diagrams in Fig.~\ref{Feynman}(c) form a gauge
invariant subset, but not for the ${}^3\!S_1^{[1]}+{}^3\!S_1^{[8]}$ channel
because of $g\to c\bar{c}({}^3\!S_1^{[8]})$ formation. 
Their leading large-$|Y|$ contributions can be estimated via the gauge
invariant MRK-asymptotic formalism, already used to evaluate
$d\sigma^{\mathrm{BFKL},0}$ for the LT subprocesses in Eq.~(\ref{tot}).
  We have checked for the
  $R^{+}R^{-}\to c\bar{c}({}^3\!S_1^{[1]})c\bar{c}({}^1\!S_0^{[8]})g$ 
  subprocess that our MRK approximation reproduces the exact result for the
  $t$-channel gluon exchange type diagrams in the upmost CMS $|Y|$ bin within a
  factor of 1.2.
 
In this way, we find that such partial NLO (NLO${}^{\ast}$) results for the
individual $(m,n)$ channels among the NLT subprocesses can be up to a hundred
times larger than the LO PRA results for these channels in the upper $|Y|$ and
$m_{\psi\psi}$ bins.
The effect of adding the total ${\rm NLO}^{\ast}$ NLT contribution on top of
the central LO NRQCD prediction in the PRA with BFKL resummation is shown for
the upmost $m_{\psi\psi}$ and $|Y|$ bins in Figs.~\ref{pracms} and
\ref{praatlas}.
In Fig.~\ref{pracms}(c), this amounts to 45\% and 16\% for direct and prompt
production, respectively.
The total ${\rm NLO}^{\ast}$ NLT contributions will, in turn, be enhanced by
BFKL resummation, which we leave for future work.

To summarize, we have pushed the NRQCD factorization approach to double prompt
$J/\psi$ hadroproduction beyond LO in two important ways.
On the one hand, we have incorporated multiple gluon radiation off the initial
state via the PRA, which, unlike other $k_T$ factorization approaches
frequently used in the literature \cite{Baranov:2015cle}, ensures for the SDCs
to be manifestly gauge invariant, infrared safe, and devoid of artificial
kinematic cuts.
On the other hand, we have resummed, via BFKL evolution in $|Y|$, the LLs of
the form $(\alpha_s\ln|s/t|)^n$ arising from $t$-channel gluon exchanges in the
LT subprocesses [see Fig.~\ref{Feynman}(b)], which would otherwise inevitably
invalidate the fixed-order treatment at large $|Y|$ and $m_{\psi\psi}$ values.
This consolidates the theoretical basis for meaningful extractions of the DPS
key parameter $\sigma_{\mathrm{eff}}$.

\begin{acknowledgments}
M.A.N. was supported by the Alexander von Humboldt Foundation through a
Research Fellowship for Postdoctoral Researchers.
V.A.S. was supported in part by Samara University Competitiveness Improvement
Program under Task No.\ 3.5093.2017/8.9.
This work was supported in part by BMBF Grant No.\ 05H18GUE and DFG Grant No.\
KN~365/12-1.

\end{acknowledgments}

\end{document}